# Angstrom-scale ion-beam engineering of ultrathin buried oxides for quantum and neuro-inspired computing

Nikita Smirnov[1,2], Elizaveta Krivko[1,2], Daria Moskaleva[1,2], Dmitry Moskalev[1],
Anastasia Solovieva[1], Vladimir Echeistov[1], Evgeny Zikiy[1], Nikita Korshakov[1],
Anton Ivanov[1], Elizaveta Malevannaya[1], Aleksei Matanin[1], Viktor Polozov[1],
Maksim Teleganov[1], Nikolai Zhitkov[1], Ruslan Romashkin[2], Igor Korobenko[1], Aleksei Yanilkin[2],
Andrey Lebedev[2], Ilya Ryzhikov[1,3], Aleksander Andriyash[2] and Ilya Rodionov[1,2*]

[1] Shukhov Labs, Quantum Park, Bauman Moscow State Technical University, Moscow, 105005, Russia
[2] Dukhov Automatics Research Institute, VNIIA, Moscow, 127030, Russia
[3] Institute for Theoretical and Applied Electrodynamics, Russian Academy of Science, Moscow, 125412, Russia
*e-mail: irodionov@bmstu.ru

## Abstract

Multilayer nanoscale systems incorporating buried ultrathin tunnel junctions, 2D materials, and solid electrolytes are crucial for next-generation logics, memory, quantum and neuro-inspired computing. Still, an ultrathin layer control at atomic scale is challenging for cutting-edge applications. Here we introduce a scalable approach utilizing focused ion-beam irradiation for buried ultrathin layers engineering with angstrom-scale thickness control. Our molecular dynamics simulations of $Ne^+$ irradiation on Al/a-AlO$_x$/Al structure confirms the pivotal role of ion generated crystal defects. We experimentally demonstrate its performance on Josephson junction tunning in the resistance range of 2 to 37% with a standard deviation of 0.86% across 25×25 mm chip. Moreover, we showcase ±17 MHz frequency control (±0.172 Å tunnel barrier thickness) for superconducting transmon qubits with coherence times up to 500 μs, which is promising for useful fault-tolerant quantum computing. This work ensures ultrathin multilayer nanosystems engineering at the ultimate scale by depth-controlled crystal defects generation.

## Main

Emerging fields of quantum computing and artificial neural networks based on ion-dynamic capacitance[1], memristors[2], single flux quantum (SFQ) logic[3], and quantum machine learning[4–6] boost the way in post-exascale hybrid data processing. Useful applications exploiting these principles require thousands or millions logical elements, whose key parameters are determined by ultrathin oxides (0.5-40 nm thick) buried within multilayer nanoscale systems [7–12]. Widespread diffusive memristor computing with large artificial neuron arrays is limited by its unstable relaxation dynamics governed by the microscopic properties of a sub-10nm oxides [7,8]. Multimode transistors and neural networks based on ion-dynamic capacitance require specific frequency- and voltage-dependent capacitance formed by a 40-nm-thick aluminum oxide [10,11]. Statistical variations in Josephson junctions (JJs) critical current (1.5-10 nm AlO$_x$[12] or a-Si[13] tunnel barriers) of SFQ circuits affect bit error rates and high frequency bias margins [14]. Superconducting quantum processors suffer from two-qubit gate errors [17] and crosstalks [18]. As the number of qubits grow up, the probability of frequency collisions increases exponentially. Recent superconducting multiqubit processors require qubit frequency set up accuracy ±0.5%, which corresponds to nanoscale JJs normal resistance variation less than 1% (tunnel barrier thickness of 0.5-2 nm) [15,19]. When scaling these devices their nanoscale building block reproducibility becomes crucial and unattainable for state-of-the-art nanotechnology [1,13–16]. Laser [15,20], e-beam [22] and alternating bias [23] annealing methods were proposed for ultrathin oxide adjusting after fabrication. Their practical applications are limited by non-local nature [15,20,23], narrow tunning range [22] or unestablished precision [22,23]. Moreover, next-generation computing platforms require local buried oxides engineering with a sub-atomic accuracy and chip-scale throughput, together with non-destructive effect on overlying layers of multilayer nanoscale systems.

In our approach, we ensure sub-Å accuracy oxide growth buried within multilayer nanoscale system by defect-induced annealing of top oxide interface with ~3 nm spot focused ion-beam (FIB). Meanwhile, it allows sub-5nm patterning by scanning multilayer nanoscale systems along a specified trajectory (a given topology). One can select the desired layer in the stack by depth, which is controlled with ion acceleration voltage, and the growth thickness determined by the radiation dose with sub-angstrom precision (Fig. 1a,c). Our physical model of the oxide ion-beam annealing beneath the top metal is based on molecular dynamics (MD) simulations of ions impact on material structure. We investigate experimentally the effect of ion radiation dose on the room temperature resistance of Josephson junctions and demonstrated its reproducibility up to 0.86% over 25×25 mm chip. We next measure a number of transmon-based quantum processors in a cryogenic environment and guarantee qubit frequency trimming accuracy $\pm 17 MHz$ using post-fabrication ion-beam annealing. Moreover, we confirmed that it does not affect superconducting qubits maintaining coherence times up to 500 μs. The proposed method can be highly effective in scaling superconducting quantum circuits (fluxonium[24,25], parametric amplifiers[26–28], quantum radars[29]), powerful neuromorphic computing networks[30–33], low-power integrated circuits, sensitive biosensors and gas sensors, thermal diodes[34], and new types of memory[35–37] (magnetic, semiconductor, etc.).

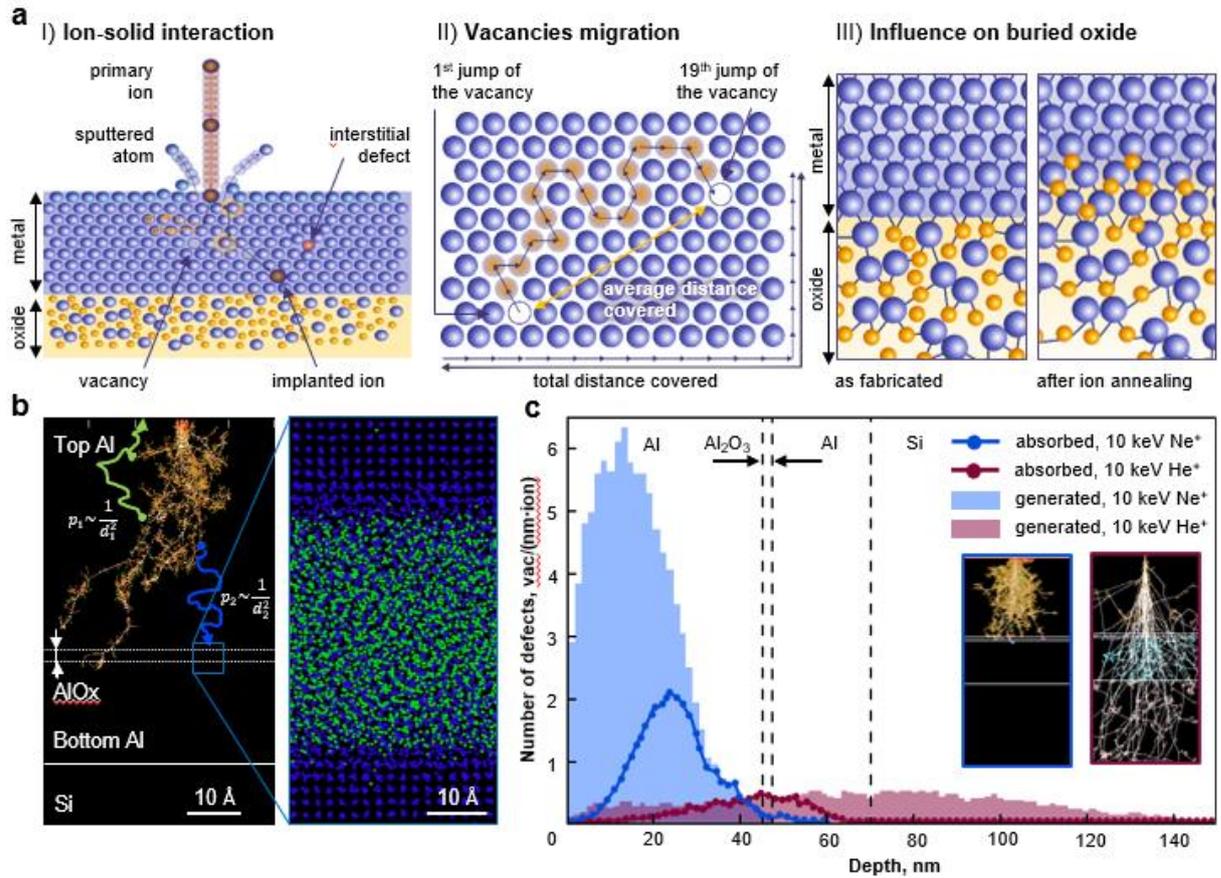

**Fig. 1: Ion-beam engineering of buried oxides: effects of ion irradiation on multilayer nanoscale systems. a,** Basic principles: (I) the ion beam interacts with the sample, creating vacancies and interstitial defects; (II) probabilistic defects diffusion towards the oxide interface; (III) crystal defect-induced oxygen diffusion toward metal-oxide interface and oxide formation. **b,** SRIM (left) and MD (right) simulations for 10 keV Ne$^+$ irradiation of Al/AlO$_x$/Al nanoscale system. Left: Ne+ create interstitial defects and vacancies, which migrate to the upper or lower interfaces with probabilities dependent on the distance to the interfaces. Right: defects and vacancies reaching the oxide interface induce oxide recrystallization. **c,** Depth distribution of the generated and absorbed defects after 10 keV Ne$^+$ and He$^+$ irradiation. Absorbed defects are calculated based on the depth distribution of generated defects (SRIM simulations): the sum of all ion ranges over the surface, top and bottom metal-oxide interfaces.

## Ion-irradiation-induced crystal defects formation

Numerical simulations using SRIM Monte Carlo approach [39,40] (Fig. 1b) and molecular dynamics simulations (Fig. 2) confirmed that oxide layers growth is driven by the diffusion of radiation-induced defects to metal-oxide interfaces. Incident ions form a collision cascade along their trajectories, leading to materials amorphization followed by rapid recrystallization. Consequently, defects are generated that slowly diffuse towards interfaces, where they subsequently relax, resulting in material restructuring along the interfaces. For example, one incident 10 keV Ne+ generate 150 defects (vacancies and interstitials) in an Al/a-AlO$_x$/Al nanoscale system, while one 10 keV He+ generate only 50 defects (Fig. 1c). At room temperature, these defects can diffuse for the distance up to 100 nanometers. According to MD simulations interstitial defects are more mobile during tens of picoseconds timescale. Vacancies have a lower diffusion coefficient (1 to 100 nm²/s), depending on the calculation method[41]. Therefore, vacancies remain immobile over microsecond timescale, but may reach interfaces over seconds to minutes. In our Al/a-AlO$_x$/Al system, mobile defects in the top aluminum layer have two relaxation pathways (Fig. 1b): migration to the surface (green trajectory) or oxide interface (blue trajectory). The probability of these processes is inversely proportional to the square of the distance to the interfaces [42]: $p_{abs} \sim \frac{1}{d^2} / \sum \frac{1}{d_i^2}$. When a defect reaches the interface, it triggers local atomic structure reconfiguration, as confirmed by MD calculations (Fig. 2b,c and Supplementary Section 2). The more defects reach the metal-oxide interface, the greater the transformations in the oxide layer. The change in tunnel Josephson junction resistance is proportional to the ion dose; however, saturation may occur beyond a certain dose. We performed MD simulations to precisely analyze the mechanism of high-energy Ne+ interaction with Al/a-AlOx/Al nanoscale system containing a buried ultrathin tunnel oxide. These simulations revealed three characteristic scenarios of ion impact and propagation (Supplementary Section 2). Besides the defect-induced growth scenario, we explored alternative scenarios [43] and concluded they affect negligible on buried oxides (Supplementary Section 4).

**Scenario 1:** The ion does not reach the oxide due to a strong collision with an aluminum atom (Fig. 2a (I)). As the result, an amorphous region and the majority of defects form in the middle of the top aluminum layer. After the amorphous region recrystallization numerous defects remain in the form of vacancies and interstitial clusters, which are mobile at room temperature over the MD simulation timescale.

**Scenario 2:** The ion reaches the oxide layer, where most of its energy is released (Fig. 2a (II)). This causes amorphization of both the top Al layer and the oxide layer. Figure 2b shows the state of the oxide layer at various times before and after the cascade, as well as post-relaxation. We hypothesize that a small amount of unbound oxygen from the oxide layer diffuses into the aluminum. However, within 5-20 picoseconds, annealing occurs, leading to aluminum recrystallization. Consequently, oxygen does not significantly diffuse into the aluminum, with only a few atoms moving beyond two lattice periods. Most changes occur within one lattice period ~4 Å. Notably, the second scenario also results in a significant number of defects in the oxide layer (Supplementary videos). Therefore, it can be concluded that displacement cascade formation in the oxide layer causes the amorphization of the adjacent aluminum layer, leading to local interface restructuring due to defect relaxation and minor oxygen displacement (up to 5-10 Å).

**Scenario 3:** The ion penetrates the entire Al/a-AlO$_x$/Al structure, stopping in the lower Al layer (Figure 2a (III)). Consequently, the amorphous region and the majority of defects form in the lower part of the aluminum layer, close to the oxide layer.

These scenarios demonstrate that even a single neon ion generates a substantial number of mobile lattice defects in the metal layers, capable to diffuse towards the oxide. Hence, the interface serves as an efficient sink for crystalline lattice defects. Figure 2c illustrates the absorption process of interstitial defects by the oxide-aluminum interface. Lattice defects at the interface lead to local structural changes, enabling angstrom-scale control over the properties and thickness of ultrathin oxides buried under thin films of other materials using ~3-10 nm spot focused ion beam irradiation. Furthermore, by varying the accelerating voltage, dose and ion type, it is possible to precisely target materials buried at the exact depth of multilayer nanoscale systems. With this experimental nature in mind, we name our method the 'iDEA' (ion-beam-induced DEfect Activation) annealing.

## Josephson junction a-AlO$_x$ engineering at atomic scale

We investigate the influence of FIB irradiation on buried oxides within multilayer nanoscale systems using extremely sensitive Al/a-AlO$_x$/Al Josephson junctions with ultrathin buried tunnel oxides (0.5-3nm-thick a-AlO$_x$) and top/bottom aluminum thin electrodes (15-100nm-thick) as a model system (Fig. 3a). The thickness and structure of a tunnel oxide determine its critical current, which is related to the normal resistance $R_N$ through Ambegaokar-Baratoff relationship [44]. We measured JJs critical current at room temperature before and after iDEA annealing. Top/bottom aluminum thickness (45/25 nm) of JJs is typical for superconducting quantum processors with the state-of-the-art JJs area variation ~3% [45,46]. See the Methods for a description of the fabrication process and test JJs structures measurement. Local FIB treatment was then applied to each JJs (Fig. 3b), avoiding the substrate area (Methods). The normal resistance was measured immediately after fabrication and iDEA annealing.

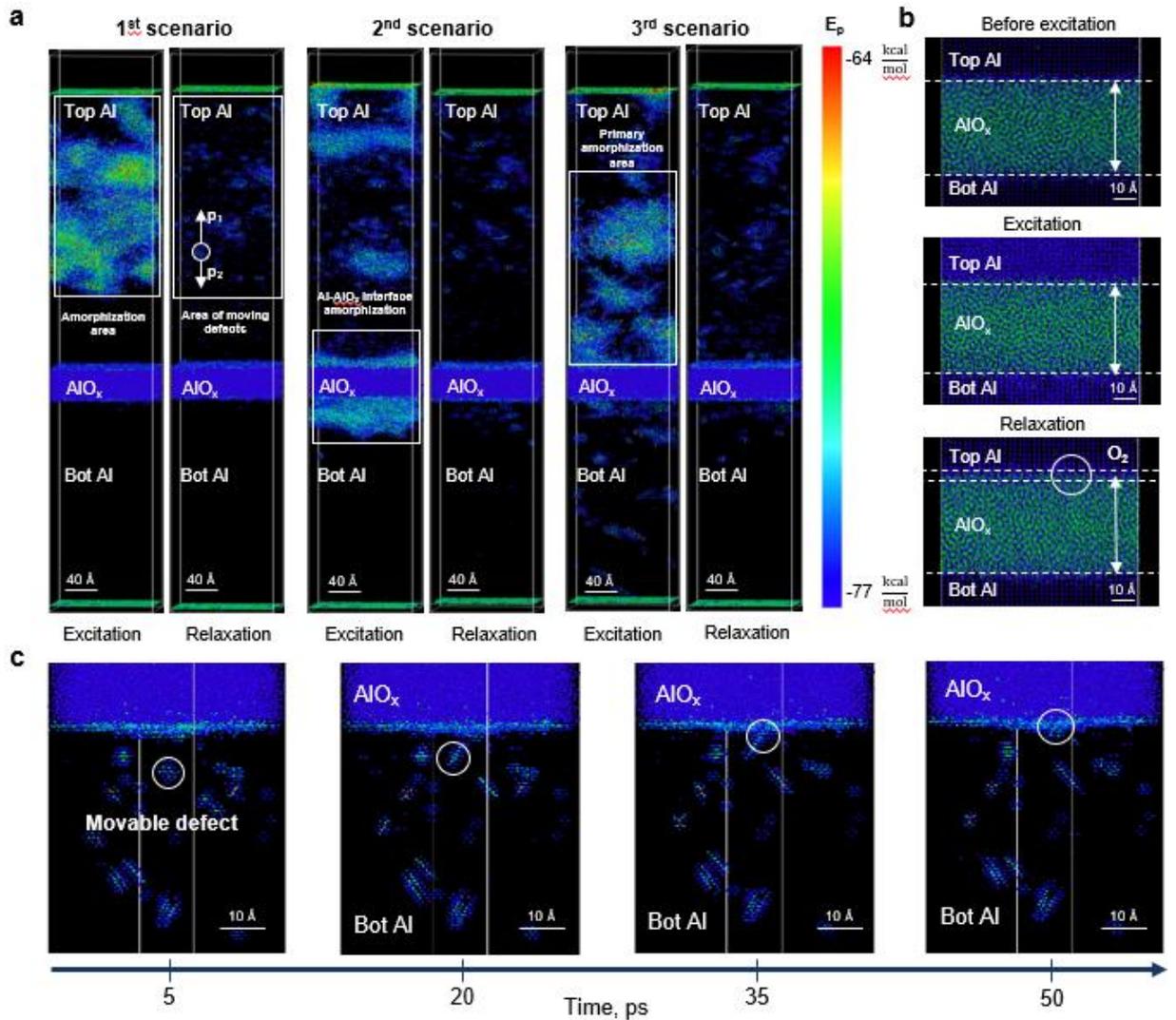

**Fig. 2: MD simulation of Ne$^+$ irradiation on Al/a-AlO$_x$/Al (37.0/3.8/19.0 nm) with 8.3×8.3 nm² cross-section and 0.5 fs time resolution. a,** Characteristic scenarios of ion interaction with the Al/a-AlO$_x$/Al structure are shown: ion deceleration in the top Al, the tunnel barrier, and the lower Al. It is evident that the amorphization region and the majority of defects form predominantly in the top Al and the tunnel oxide in all cases. **b,** The oxide and its interface with the top Al at the initial moment, after the displacement cascade, and post-relaxation. Oxide changes are observed in the Al-AlO$_x$ interface after the relaxation of ion-induced defects. **c,** Interstitial defect absorption at successive time points: the interstitial defect approaching the interface and its subsequent absorption.

We experimentally determined the dependence of JJs normal resistance on the irradiation dose of 10 keV Ne$^+$ at room temperature (Fig. 3d). The irradiation dose range was chosen to eliminate the sputtering of aluminum atoms (from $0.015×10^{13}$ to $8×10^{13}$ ions/cm²), which was confirmed by SRIM calculations and scanning electron microscopy. For each irradiation dose 25 identical JJs were annealed and measured (Fig. 3d). We enable increasing the room-temperature resistance over the wide range of 2 to 37% in a controllable way using Ne$^+$, and of 2% to 20% with He$^+$. The smaller effect of He$^+$ on the buried oxide layer is due to their lower mass and deeper penetration depth (Supplementary Section 1,2). The relative resistance variation is independent on the junction area and is determined solely by the irradiation dose (i.e., the number of defects absorbed by the metal-oxide interface). This dependence is well approximated by a linear function (Fig. 3d) allowing post-fabrication JJs critical current tuning with extreme accuracy. Finally, we have proved no tunnel barrier crystalline structure degradation (Methods, TEM) for the JJs with the highest irradiation dose for Ne+ ($8\times10^{13}$ ions/cm² at 10 keV) and 36% normal resistance increase after iDEA annealing (Fig. 3c).

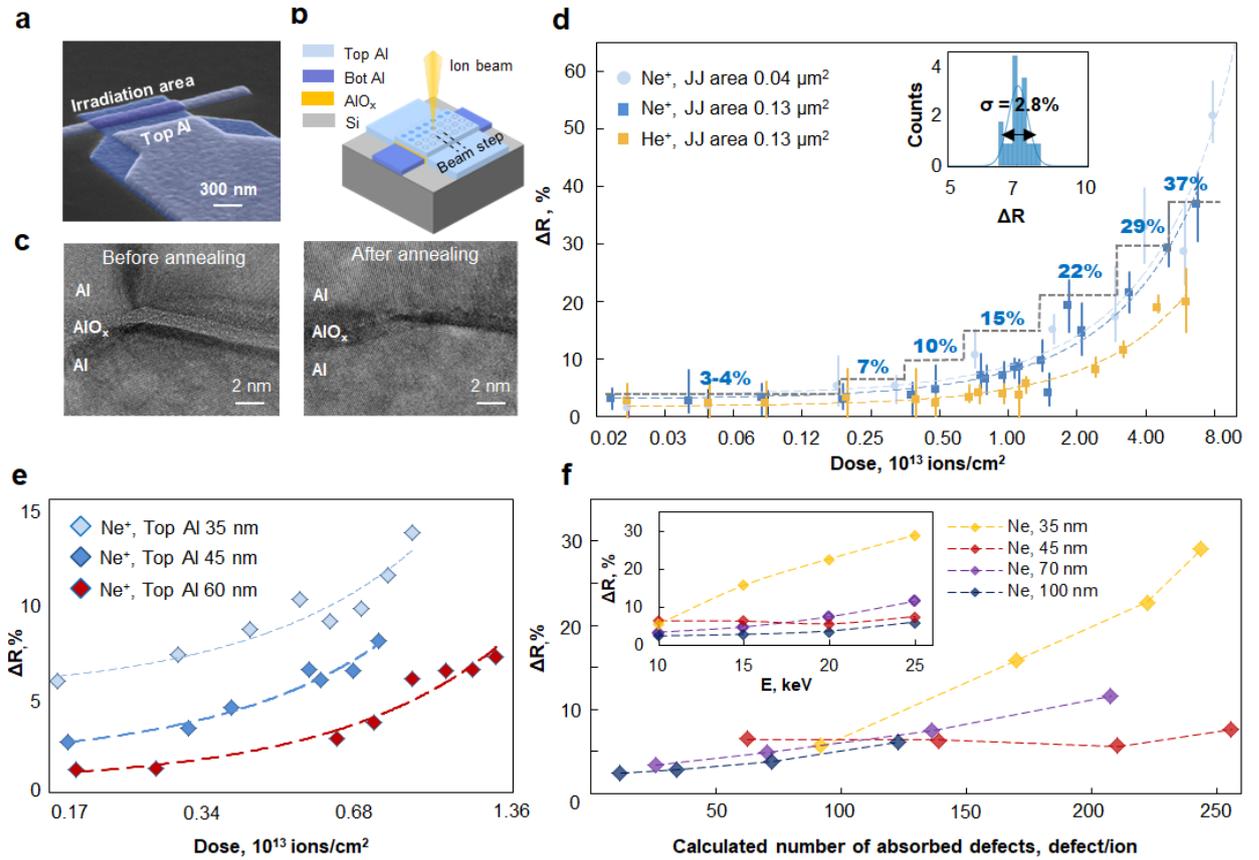

**Fig. 3: Josephson junctions iDEA annealing (areas 0.04 μm² and 0.13 μm²). a,** Typical Al/a-AlOₓ/Al tunnel junctions test structure. **b,** Scanning focused ion-beam annealing of Josephson junctions with 10 nm spot step (down to 5 Å). **c,** TEM images of characteristic top/bottom Al crystallites and tunnel barrier before (left) and after (right) iDEA annealing. No changes in the tunnel barrier crystalline structure are detected. **d,** Experimental curve (logarithmic scale) of room-temperature JJs resistance on 10 keV He+ and Ne+ irradiation dose. The range of resistance variation exceeds 50%, the accuracy is independent on junction area and determined solely by irradiation dose. **e,** Experimental curve of resistance variation on 10 keV Ne+ dose for different top Al layer thickness. Top electrode thickness increasing up to 60 nm reduces the effect of ion irradiation on the tunnel barrier, allowing a tighter resistance control over a wider range of processing doses. **f,** Experimental curve of resistance variation on the number of absorbed defects for different top Al layer thickness and Ne+ accelerating voltages.

Next, we studied the effect of Ne+ irradiation with various top Al thicknesses (Fig. 3e) and accelerating voltage (Fig. 3f) to confirm the proposed hypothesis of iDEA annealing mechanism and sub-Å thickness control potential. Decreasing the top Al electrode thickness (constant 10 keV and penetration depth), the effect of Ne+ on the tunnel barrier increases (Fig. 2a) due to the reduced distances to the metal-oxide interface and the increased probability of defect absorption by the interface. Next, we determined the dependence of relative resistance variation on the number of absorbed defects. The impact on the oxide was quantified by the number of defects absorbed by top (metal-oxide) and bottom (oxide-metal) interfaces. We found the number of absorbed defects (SRIM simulations) for the experiments with different top Al thickness (35, 45, 70, and 100 nm), various Ne+ irradiation (10, 15, 20, and 25 keV) at a fixed dose of $1\times10^{13}$ ions/cm². Experimentally, we confirmed the increase in tunnel oxide resistance with increased number of absorbed defects (Fig. 3f).

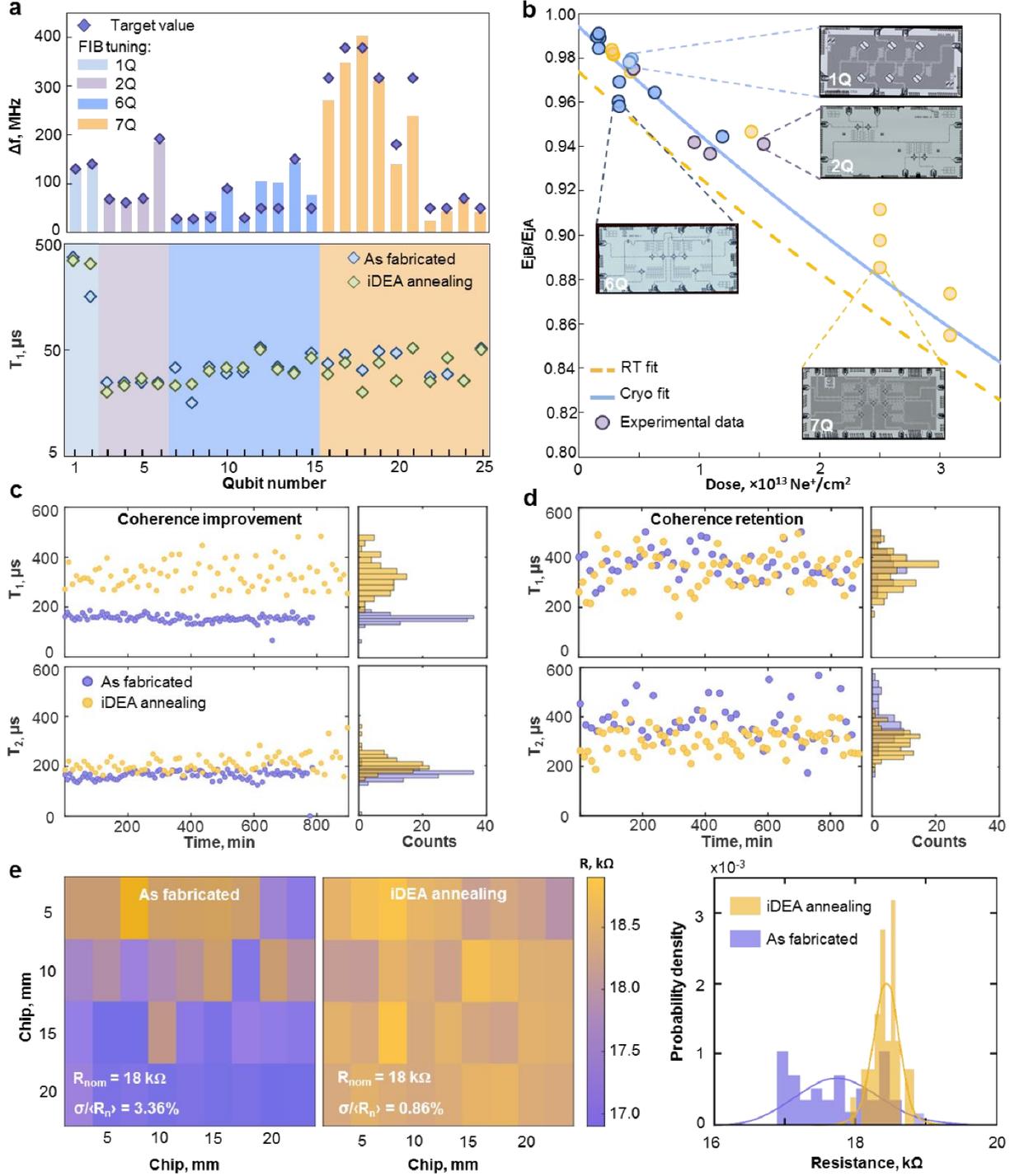

**Fig. 4: iDEA annealing of superconducting trnamons and multiqubit quantum processors. a,** Experimental results on the frequency trimming accuracy for superconducting transmon qubits and iDEA annealing effect on qubits relaxation time ($T_1$). Different qubit circuits represented in colors: single-qubit (blue), two-qubit (purple), six-qubit (blue), and seven-qubit (yellow). Overall frequency trimming accuracy is $\sigma_f = 17 MHz$. **b,** Correlation between transmon Josephson energy shift and irradiation dose based on room-temperature resistance measurements (dashed yellow line) and qubits cryogenic characterization (blue). The correlation coefficient was 0.78 (constant shift of JJs resistance at room and cryogenic temperatures). The factor of 1.02 is used to account this effect when calculating the Josephson energy with Ambegaokar-Baratoff formula. **c-d,** Time-domain measurements of highly coherent qubits before and after iDEA annealing. Both the qubit coherence improvement from 150 up to 400 μs (c) and retention up to 500 μs (d) are observed after iDEA annealing. **e,** Distribution maps and probability density histograms of normal resistance $R_N$ for 0.03 μm² Josephson junctions across 25x25 mm chip before and after iDEA annealing.

## Superconducting multiqubit processor frequency trimming

We also evaluated the performance of iDEA annealing in post-fabrication frequency trimming of superconducting single transmons, two-, six-, and seven-qubits quantum processors (chips) through nanoscale treatment of individual Josephson junctions. This step is crucial to eliminate unwanted qubit frequency collisions and detuning between interacting qubits for fault-tolerant quantum computing. First, we measured the transition frequencies $\omega_{01}$ and $\omega_{12}$ between energy levels $|0\rangle \rightarrow |1\rangle$ and $|1\rangle \rightarrow |2\rangle$, respectively (Methods, Cryogenic setup). A transmon[47] charging energy $E_C$ can be determined with the difference in measured transition frequencies: $E_C = \hbar(\omega_{01} - \omega_{12})$, which, together with $\omega_{01}$, allows determining $E_J$ and, thus, the normal resistance $R_N$ using Ambegaokar-Baratoff relationship. After warming up the chips, we irradiated each JJ of the qubits to get the desired $E_J$ for the target qubit frequency, using the experimental curve (Fig. 4b). Then, we cooled the chips again and measured frequencies of 25 qubits in different qubit circuits (Fig. 4a). In the case of 7-qubit circuit (Fig. 4a, yellow), we also tested a two-step iDEA annealing: after the first cool down, the qubits frequency was trimmed by 300-400 MHz, followed by precise 50-70 MHz tuning after the second cool down.

For all the 25 qubits we measured relaxation times before and after iDEA annealing (Fig. 4a). For two highly coherent qubits[48] relaxation ($T_1$) and coherence ($T_2$) in time were investigated. One can note, a stable increase of relaxation from 150μs up to 400μs (Fig. 4c), while the coherence remained unchanged. We hypothesize that this improvement is due to the detuning of the qubit out of strongly coupled two-level system[49–51] after annealing. The average change in relaxation time after iDEA annealing was about 3%. We concluded, that iDEA annealing does not adversely affect the coherence of extremely sensitive superconducting quantum multilayer nanoscale systems (Fig. 4). Notably, for frequency trimming above 150 MHz the accuracy decreases, however, a multi-step iDEA annealing can be used for better precision. Finally, we achieved a frequency trimming accuracy of $\sigma_f = 17 MHz$ (Fig. 4d).

Moreover, we have improved the chip-scale variation coefficient of room-temperature resistance $\sigma_{R_N}/\langle R_N \rangle$ (JJs nominal resistance of 18 kΩ and area of 0.03 $\mu m^2$) from 3.36 (as fabricated) to 0.86% (corresponding to a critical current variation of 0.87%), which is the best result for JJs yield at the same substrate area[19,52–56] (Supplementary Section 3). Assuming the typical 20 Å initial thickness of the tunnel oxide, for the experimentally measured standard deviation of room temperature resistance of ±0.86%, one can estimate ±0.172 Å accuracy of tunnel oxide thickness across the substrate area.

## Conclusions

In summary, we introduce a novel approach for nanoscale control of buried oxides inside multilayer nanoscale systems with sub-Å precision via focused ion-beam irradiation. It involves creating localized crystal defects in the overlying layer above the target oxide using incident ions, which diffuse thought top stack layers in a non-damage manner towards depth-selected metal-oxide interface. One can select the desired layer in multilayer stack by depth, which is controlled with ion acceleration voltage, and the growth thickness determined by the radiation dose with atomic precision. Molecular dynamics simulations demonstrate that single ions excite numerous mobile defects capable diffusing to the oxide boundary and inducing local structural changes.

We experimentally validated iDEA approach using sensitive Al/a-AlOx/Al Josephson junctions and quantum nanoscale systems with ultrathin buried tunnel oxide as a model. We substantiated the proposed annealing mechanism by quantitatively analyzing resistance change as a function of absorbed oxide layer defects. The method enabled controlled adjustment of the tunnel junction normal resistance within the range of 2% to 50% without structural damage, and the relative resistance variation remained independent of junction area. We utilized it to improve the chip-scale normal resistance variation of Josephson junctions from 3.36% to 0.86% (a critical current variation of 0.87%). The achieved variation is the best reported for as-fabricated Josephson junctions of the same nominal resistance and area (18 kΩ and 0.03 $\mu m^2$ respectively). We have confirmed the performance of iDEA annealing for post-fabrication frequency trimmimg of superconducting single transmons, two-, six-, and seven-qubits quantum processors, achieving adjustments accuracy down to 0.5%. Importantly, high qubit coherence was preserved, with some instances showing a coherence increase from 150 up to 400 μs. This approach provides new capabilities of angstrom-scale control of multilayer nanoscale system for post-exascale hybrid computing.

## Methods

### Nanoscale ion-beam engineering of buried oxide

A primary ion beam with specified energy and current was focused onto the sample, determining the contact spot diameter by selecting the aperture and beam current. Subsequently, the selected area on the sample was scanned using deflecting coils, with scan step and direction controlled by the user. In this study, the scanning algorithm adhered to a fixed topology trajectory resembling a meander pattern. The mechanized stage then moved to the next structure. Processing time for each structure (up to 1 second) depended on the area size under treatment.

The processing dose $Dose = \frac{N}{A} = \frac{I \cdot t}{q \cdot A}$ was defined by the primary ion beam current and exposure time t, where N represents the number of ions penetrating per unit area A, I is the focused ion beam current, t is the exposure duration, and q denotes the ion charge. The primary beam current was determined by the diameter of the aperture and the working gas pressure ($2 \times 10^{-6}$ mbar).

### Molecular dynamics simulation

We employed a material stack replicating Josephson junctions for our simulations. A computational cell comprising an upper aluminum layer (37 nm thick), an aluminum oxide layer (3.8 nm thick), and a lower aluminum layer (19 nm thick) was created. The lateral dimensions were set at 8.3 nm. ReaxFF interatomic potential with short range part ZBL is used to describe the interactions between atoms. We performed the feasibility analysis of collision cascade in aluminum using MD to check the ReaxFF potential (Supplementary Section 2). The oxide layer was initially amorphized through MD simulations at high temperature, followed by relaxation at 300 K. Subsequently, the oxide layer was interfaced with the aluminum layers. During the relaxation process, the aluminum-oxide boundary formed crystal lattice defects, with some oxygen migrating into the upper aluminum layer, creating a transition layer [Fig. 2a(I)].

Following structural relaxation, sequential ion irradiation with neon ions was conducted above the upper aluminum layer, each ion imparted a velocity corresponding to 30 keV energy. Each ion irradiation calculation involved two stages: the first stage simulated the passage of a high-energy ion and the creation of a displacement cascade, while the second stage encompassed cascade relaxation, thermalization, recrystallization, and diffusion processes. The first stage simulations were performed with a variable integration step to resolve high-energy collisions. The second stage employed a fixed 0.5 fs integration step over 100,000 steps, equivalent to a total simulation time of 50 ps. Thermalization via interaction with the remaining materials was simulated using thermostating, adjusting the temperature to room temperature over several picoseconds.

### Josephson junctions test crystal and fabrication

For this study, we used high-resistivity silicon substrates (10,000 Ω cm). Prior to the base layer deposition, the substrate is cleaned in a Piranha solution at 80 °C, followed by dipping in hydrofluoric bath. 100 nm thick Al base layer is deposited using ultra high vacuum e-beam evaporation system. Pads were defined using a direct-laser lithography and dry-etched in $BCl_3/Cl_2$ inductively coupled plasma. The Josephson junctions were fabricated using Dolan technique. The substrate is spin coated with resist bilayer composed of 500 nm EL9 copolymer and 100 nm CSAR 62. Layouts were generated and exposed with 50 keV e-beam lithography system. The development was performed in a bath of amylacetate followed by IPA dip and additional in a IPA-DIW solution. Al/AlO$_x$/Al junctions are shadow evaporated in ultra-high vacuum deposition system. Resist lift-off was performed in N-methyl-2-pyrrolidone at 80 °C. Then aluminum bandages are defined and evaporated using the same process as for the junctions with an in-situ Ar ion-milling to provide good electrical contact of the junction with the base layer. Lift-off is performed in a bath of N-methyl-2-pyrrolidone with sonication at 80°C and rinsed in a bath of IPA with ultrasonication. The room temperature resistances of Josephson junctions were individually measured after fabrication and then after ion irradiation with automated probe station.

**TEM, SEM & EDS characterization of a-AlO$_x$ tunnel barriers**

During the fabrication of test and experimental samples, scanning electron microscopy was employed for quality and uniformity assessment of deposited Josephson junction electrodes. Transmission electron microscope (TEM) samples were prepared by a focused ion beam instrument with a gas injection system. The TEM samples were thinned to electron beam transparency by a Ga+ ion beam from 30 to 2 kV. The TEM samples were investigated by an aberration-corrected TEM at 200 kV. A high-angle annular dark-field detector was used for dark-field imaging in scanning TEM mode with a convergent semi-angle and a collection semi-angle of 18 m rad and 74-200 m rad, respectively. Energy-dispersive X-ray spectroscopy (EDS) studies were carried out with probe currents of 250 nA.

**Room temperature normal resistance characterization**

The Josephson junctions room temperature resistance were individually measured using standard technique based on passing current through them and measure the voltage drop across the same junction with automated probe station. Each test chip consisted of an array of 1600 Josephson structures (Supplementary Section 3), and in total 12536 such structures were measured. Each test structure consisted of a single Al/a-AlO$_x$/Al junction and aluminum contact pads for electrical measurements. Multiple measurements of the same junction have demonstrated repeatability of measurements better than 0.5%.

**Multiqubit processors fabrication**

The device is made in a four-step process: (I) Base Al layer patterning, (II) Josephson junction double-angle evaporation and lift-off, (III) patterning and deposition of bandages, (IV) air-bridges fabrication[57]. Devices are fabricated on Topsil Global Wafers high-resistivity silicon substrate (ρ > 10000 Ohm · cm, 525 μm thick). Prior to the deposition the substrate is cleaned in a Piranha solution at 80°C, followed by dipping in 2% hydrofluoric bath to remove the native oxide. 100 nm thick base aluminum layer is grown using e-beam evaporation in a ultra-high vacuum deposition system. 600 nm thick Dow MEGAPOSIT SPR 955-CM photoresist is then spin coated. Qubit capacitors, resonators, wiring and ground plane are defined using a laser direct-writing lithography system, developed in AZ Developer to minimize film damage and then dry etched in BCl3/Cl2 inductively coupled plasma. The photoresist is stripped in N-methyl2-pyrrolidone at 80°C for 3 h and rinsed in IPA (isopropyl alcohol) with sonication. The substrate is then spin coated with resist bilayer composed of 500 nm MMA (methyl methacrylate) and 300 nm PMMA (poly methyl methacrylate). The development is performed in a bath of MIBK/IPA 1:3 solution followed by rinse in IPA. Al/AlO$_x$/Al Josephson junctions are patterned using 50 keV electron beam lithography system and aluminum electrodes are shadow-evaporated in ultra-high vacuum deposition system. 25 nm thick first Al junction electrode is oxidized at 5 mbar to form the tunnel barrier and next the 45 nm thick counter-electrode is evaporated. We then pattern and evaporate aluminum bandages using the same process as for junctions with an in-situ Ar ion milling in order to provide good electrical contact of the junction with the base layer. Lift-off is performed in a bath of N-methyl-2-pyrrolidone with sonication at 80°C for 3 h and rinsed in a bath of IPA with sonication. Finally, aluminum free-standing crossovers are fabricated using conventional approach. SPR 220 3 um photoresist is spin coated and then the sacrificial layer is patterned using a direct laser writing system. The development is performed in AZ Developer / deionized water solution (1:1) for 2 minutes in order to minimize film damaging and the resist is reflowed at 140°C. 300 nm of Al is then evaporated with an in-situ Ar ion milling to remove the native oxide. Second layer of 3 um SPR 220 is used as a protective mask and the excess metal in dry etched in inductively coupled plasma. Damaged layer of photoresist is then removed in oxygen plasma and both layers of photoresist are stripped N-methyl-2-pyrrolidone at 80°C.

# Cryogenic setup

The detailed experimental setup scheme is shown in Fig. 5. The device is measured in a Bluefors LD400 dilution refrigerator. One line connected to the chip is used for readout and the others for applying single qubit gates (XY controls). Pulsed XY control of the qubits was realized by upconverting the intermediate-frequency in-phase and quadrature signals from the arbitrary waveform generator (AWG), using IQ-mixer and microwave local oscillator.

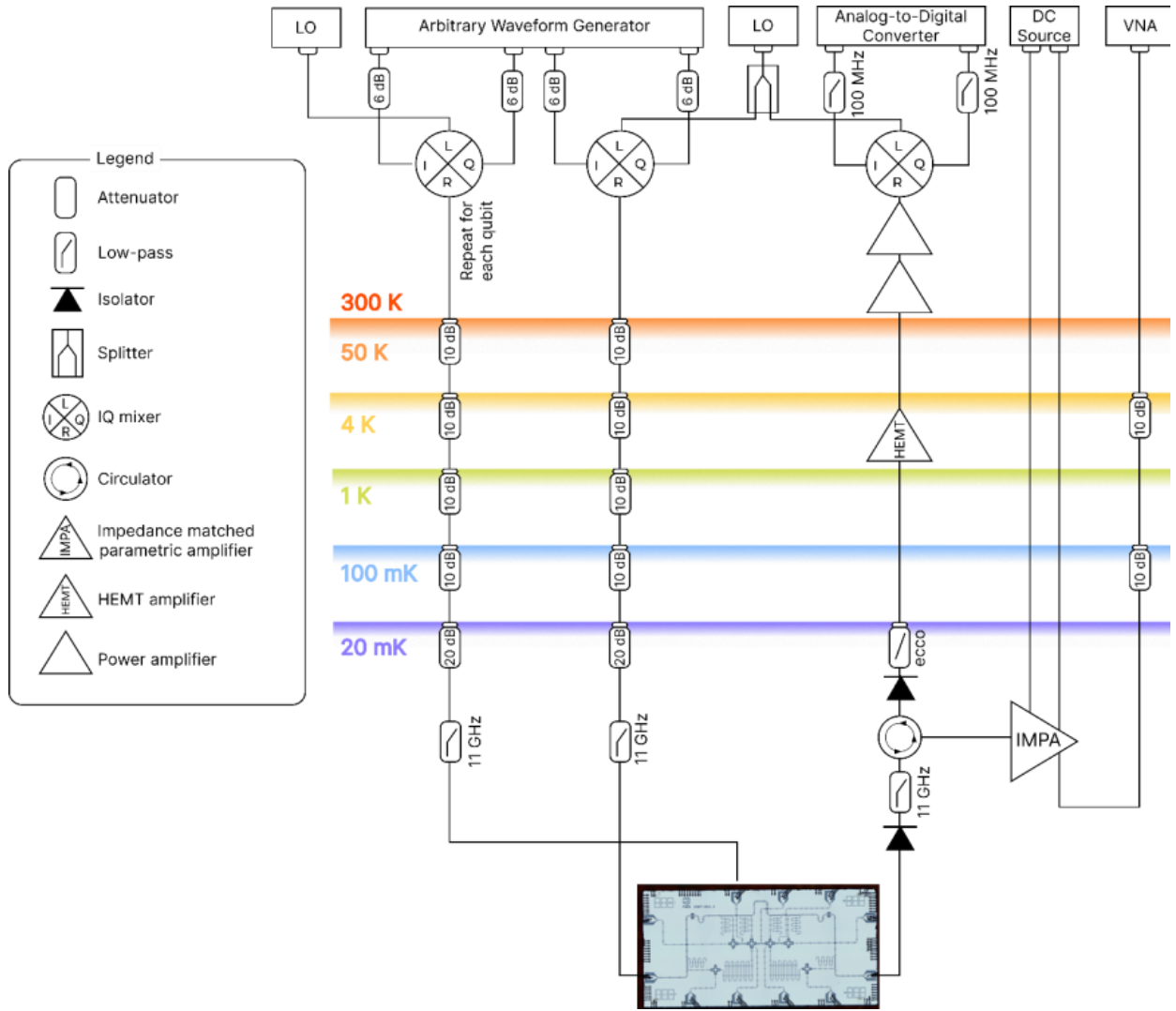

**Figure 5.** Measurement scheme of superconducting quantum processors in a dilution refrigerator.

Readout tone was generated by AWG and up-converted to the readout resonator frequency using mixer and microwave local oscillator. Readout microwave signal passed through the chip, is amplified by a cryogenic impedance matched parametric amplifier[28] (IMPA) and then downconverted. The readout signal is also amplified by high-electron mobility transistors (HEMT) at the 4K stage of the cryostat and at room temperature. We use a DC source and superconducting coil located on the packaging, to tune-up the parametric amplifier to the desired frequency and pump the IMPA by microwave source of vector network analyzer (VNA). Readout line is additionally equipped with custom-made Eccosorb filter[58] on the cryostat mixing stage to suppress IR-noise and standing waves. Sample holders with IMPA is placed in the magnetic shield.

# Data availability
The data used in this Article are available from the corresponding authors upon reasonable request.

## Acknowledgements

Technology was developed and samples were fabricated at Quantum Park (BMSTU Nanofabrication Facility, Shukhov Labs, FMNS REC, ID 74300).

## Author information

These authors contributed equally: Nikita S. Smirnov, Daria A. Moskaleva, Elizaveta A. Krivko, Dmitry O. Moskalev

## Contributions

I.R. (Ilya A. Rodionov) and S.N.S. conceived the iDEA method. S.N.S., M.A.R., K.E.A., M.D.A., L.A.V., and I.R. designed the samples. Y.A.V. performed molecular dynamic simulation with input from S.N.S., M.D.A., K.E.A., and R.I. M.D.A., M.D.O., S.A.A., K.N.D., T.M.I., S.N.S., R.I.A. (Ilya A. Ryzhikov) and I.R. developed the superconducting circuits technology and fabricated the samples. I.A.I., M.E.I., M.A.R., E.V.V., P.V.I., R.R.V., K.I.S., Z.N.M. and L.A.V. developed experimental setup and characterized the devices. K.E.A., S.N.S., M.D.A., Y.A.V., R.I.A., A.A.V. and I.R. analyzed the data. K.E.A., M.D.A., Y.A.V., L.A.V., S.N.S. and I.R. wrote the manuscript with comments from the other co-authors. A.A.V. and I.R. supervised the project and interpreted data. All authors discussed the results and contributed to the manuscript.

## Corresponding authors

Correspondence to Ilya A. Rodionov (irodionov@bmstu.ru)

## Ethics declarations

Competing interests

The authors declare no competing financial interests.